\documentclass[twocolumn,amsmath,amssymb]{revtex4}

\usepackage{graphicx}
\usepackage{dcolumn}
\usepackage{bm}

\begin{document}

\title{Nuclear structure calculations for two-neutrino double-$\beta$ decay}
\author{P.~Sarriguren$^1$, O.~Moreno$^2$, E.~Moya~de~Guerra$^2$}
\affiliation{$^1$ Instituto de Estructura de la Materia, IEM-CSIC, Serrano
123, E-28006 Madrid, Spain} 
\affiliation{$^2$Dpto. F\'isica At\'omica, Molecular y Nuclear, Univ. 
Complutense de Madrid, E-28040 Madrid, Spain}

\date{\today}

\begin{abstract}
We study the two-neutrino double-$\beta$ decay in $^{76}$Ge, $^{116}$Cd, $^{128}$Te, 
$^{130}$Te, and $^{150}$Nd, as well as the two Gamow-Teller branches that connect 
the double-$\beta$ decay partners with the states in the intermediate nuclei. 
We use a theoretical microscopic approach based on a deformed selfconsistent
mean field with Skyrme interactions including pairing and spin-isospin residual
forces, which are treated in a proton-neutron quasiparticle random-phase 
approximation. We compare our results for Gamow-Teller strength distributions 
with experimental information obtained from charge-exchange reactions. We also 
compare our results for the two-neutrino double-$\beta$ decay nuclear matrix 
elements with those extracted from the measured half-lives. Both single-state 
and low-lying-state dominance hypotheses are analyzed theoretically and 
experimentally making use of recent data from charge-exchange reactions and 
$\beta$ decay of the intermediate nuclei. 

\end{abstract}


\maketitle

\section{Introduction\label{sec:introduction}}

Double-$\beta$ decay is currently one of the most studied processes both 
theoretically and experimentally \cite{2breview1,2breview2,2breview3,ahep1,ahep2}. 
It is a 
rare weak-interaction process of second-order taking place in cases where single 
$\beta$ decay is energetically forbidden or strongly suppressed. It has a deep
impact in neutrino physics because the neutrino properties are directly involved 
in the neutrinoless mode of the decay ($0\nu\beta\beta$) 
\cite{ahep3,ahep4,ahep5}. This decay mode, not 
yet observed, violates lepton-number conservation and its existence would be an 
evidence of the Majorana nature of the neutrino, providing a measurement of its 
absolute mass scale. Obviously, to extract a reliable estimate of the neutrino 
mass, the nuclear structure component of the process must be determined accurately. 
On the other hand, the double-$\beta$ decay with emission of two neutrinos
($2\nu\beta\beta$) is perfectly allowed by the Standard Model and it has been 
observed experimentally in several nuclei with typical half-lives of $10^{19-21}$ 
years (see Ref. \cite{barabash15} for a review). Thus, to test the reliability 
of the nuclear structure calculations involved in the $0\nu\beta\beta$ process, 
one checks first the ability of the nuclear models to reproduce the experimental 
information available about the measured half-lives for the $2\nu\beta\beta$ 
process. Although the nuclear matrix elements (NME) involved in both processes 
are not the same, they exhibit some similarities. In particular, the two 
processes connect the same initial and final nuclear ground states and share 
common intermediate $J^\pi =1^+$ states. Therefore, reproducing the $2\nu\beta\beta$ 
NMEs is a requirement for any nuclear structure model aiming to describe the 
neutrinoless mode.

Different theoretical approaches have been used in the past to study the  
$2\nu\beta\beta$ NMEs. Most of them belong to the categories of the 
interacting shell model \cite{caurier05,menendez09,caurier12}, 
proton-neutron quasiparticle random-phase approximation (QRPA) 
\cite{2breview1,2breview2,2bqrpa,cheoun93,raduta93,suhonen12,faessler12,simkovic04,
alvarez04,raduta04,yousef09,mustonen13}, projected Hartree-Fock-Bogoliubov 
\cite{chandra05,singh07,tomas10}, and interacting boson model 
\cite{barea09,barea13,yoshida13}.

In this work we focus on the QRPA type of calculations. Most of these calculations 
were based originally on a spherical formalism, but the fact that some of the 
double-$\beta$-decay nuclei are deformed, makes it compulsory to deal with deformed 
QRPA formalisms \cite{simkovic04,alvarez04,raduta04,yousef09,mustonen13}. 
This is particularly the case of $^{150}$Nd ($^{150}$Sm) that has received increasing 
attention in the last years because of the large phase-space factor and relatively 
short half-life, as well as for the large $Q_{\beta\beta}$ energy that will reduce 
the background contamination. $^{150}$Nd is currently considered as one of the
best candidates to search for the $0\nu\beta\beta$ decay in the planned experiments 
SNO+, SuperNEMO, and DCBA. 

The experimental information to constrain the calculations is not limited to the 
$2\nu\beta\beta$ NMEs extracted from the measured half-lives. We have also 
experimental information on the Gamow-Teller (GT) strength distributions of the 
single branches connecting the initial and final ground states with all the 
$J^\pi =1^+$ states in the intermediate nucleus. The GT strength distributions have 
been measured in both directions from (p,n) and (n,p) charge-exchange reactions 
(CER) and more recently, from high resolution reactions, such as $(d,^2$He),  
$(^3$He,t), and (t,$^3$He) that allow us to explore in detail the low energy 
structure of the GT nuclear response in double-$\beta$-decay partners 
\cite{fujita11,frekers13,madey89,helmer97,rakers05,grewe08,guess11,thies12,
sasano12,puppe12}. In some instances there is also experimental information on 
the $log(ft)$ values of the decay of the intermediate nuclei.

Nuclear structure calculations are also constrained by the experimental 
occupation probabilities of neutrons and protons of the relevant 
single-particle levels involved in the double-$\beta$-decay process.
In particular, the occupation probabilities of the valence shells 
$1p_{3/2},1p_{1/2},0f_{5/2}$, and $0g_{9/2}$ for neutrons in $^{76}$Ge 
and for protons in $^{76}$Se have been measured in Refs. \cite{sch08} 
and \cite{kay09}, respectively. The implications of these measurements on
the double-$\beta$ decay NMEs have been studied in Refs.
\cite{suh08,simkovic09,menendez09a,moreno10}.

In this paper we explore the possibility to describe all the experimental 
information available on the GT nuclear response within a formalism based on 
a deformed QRPA approach built on top of a deformed selfconsistent Skyrme 
Hartree-Fock calculation \cite{sarri12,sarri13,sarri15}. This information 
includes global properties about the GT resonance, such as its location
and total strength, a more detailed description of the low-lying excitations, 
and $2\nu\beta\beta$-decay NMEs. The study includes the decays 
$^{76}$Ge~$\rightarrow ^{76}$Se, $^{116}$Cd~$\rightarrow ^{116}$Sn, 
$^{128}$Te~$\rightarrow ^{128}$Xe, $^{130}$Te~$\rightarrow ^{130}$Xe, and 
$^{150}$Nd~$\rightarrow ^{150}$Sm. This selection is motivated by recent 
high-resolution CER experiments performed for $^{76}$Ge($^3$He,t)$^{76}$As 
\cite{thies12},  $^{76}$Se(d,$^2$He)$^{76}$As \cite{grewe08},  
$^{128,130}$Te($^3$He,t)$^{128,130}$I \cite{puppe12}, $^{116}$Cd(p,n)$^{116}$In and 
$^{116}$Sn(n,p)$^{116}$In \cite{sasano12}, as well as for 
$^{150}$Nd$(^3$He,t)$^{150}$Pm and  $^{150}$Sm(t,$^3$He)$^{150}$Pm \cite{guess11}. 
We also discuss on these examples the validity of the single-state dominance (SSD) 
hypothesis \cite{abad84} and the extended low-lying-state dominance (LLSD) that 
includes the contribution of the low-lying excited states in the intermediate 
nuclei to account for the double-$\beta$-decay rates.

The paper is organized as follows: In Section \ref{sec:theory}, we present a
short introduction to the theoretical approach used in this work to describe 
the energy distribution of the GT strength. We also present the basic expressions
of the $2\nu\beta\beta$-decay. In Section \ref{sec:results} we present the results 
obtained from our approach, which are compared with the experimental data available. 
Section \ref{sec:conclusions} contains a summary and the main conclusions.

\section{Theoretical approach}
\label{sec:theory}

The description of the deformed QRPA approach used in this work is given 
elsewhere \cite{alvarez04,sarriguren,sarri_pp}. Here we give only a summary 
of the method. We start from a selfconsistent deformed Hartree-Fock (HF) 
calculation with density-dependent two-body Skyrme interactions.
Time reversal symmetry and axial deformation are assumed in the calculations 
\cite{vautherin}. Most of the 
results in this work are performed with the Skyrme force SLy4 \cite{chabanat98}, 
which is one of the most widely used and successful interactions. Results from 
other Skyrme interactions have been studied elsewhere 
\cite{sarri03,sarri12,sarri13,sarri15} to check the sensitivity of the GT nuclear 
response to the two-body effective interaction.

In our approach, we expand the single-particle wave functions in terms of an 
axially symmetric harmonic oscillator basis in cylindrical coordinates, 
using twelve major shells. This amounts to a basis size of 364, the total number
of independent $(N,\ n_z,\ \lambda ,\ \Omega>0)$ deformed H.O. states.
Pairing is included in BCS approximation by solving
the corresponding BCS equations for protons and neutrons after each HF iteration. 
Fixed pairing gap parameters are determined from the experimental mass differences 
between even and odd nuclei. Besides the selfconsistent HF+BCS solution, 
we also explore the energy curves, that is, the energy as a function of 
the quadrupole deformation $\beta_2$, which are obtained from constrained HF+BCS 
calculations.

The energy curves corresponding to the nuclei studied can be found in Refs.
\cite{sarri13,sarri15,sarri03}. 
The profiles of the energy curves for $^{76}$Ge and $^{76}$Se exhibit two shallow 
local minima in the prolate and oblate sectors. These minima are separated by 
relatively low energy barriers of about 1 MeV. The 
equilibrium deformation corresponds to $\beta_2 = 0.14$ in $^{76}$Ge and  
$\beta_2 = 0.17$ in $^{76}$Se. We get soft profiles for $^{116}$Cd with a 
minimum at $\beta_2  = 0.25$ and  an almost flat curve in $^{116}$Sn between  
$\beta_2  = -0.15$ and $\beta_2  = 0.25$.
We obtain almost spherical configurations in the ground states of $^{128}$Te
and $^{130}$Te. The energies differ less than 300 keV between quadrupole 
deformations $\beta_2=-0.05$ and $\beta_2=0.1$. On the other hand, for $^{128}$Xe 
and $^{130}$Xe we get in both cases two energy minima corresponding to prolate 
and oblate shapes, differing by less than 1 MeV, with an energy barrier of about 
2 MeV. The ground states correspond in both cases to the prolate shapes with 
deformations around $\beta_2=0.15$. For $^{150}$Nd and $^{150}$Sm we obtain two 
energy minima, oblate and prolate, but with clear prolate ground states in both 
cases at  $\beta_2=0.30$ and $\beta_2=0.25$, respectively. We obtain comparable 
results with other Skyrme forces. The relative energies between the various 
minima can change somewhat for different Skyrme forces 
\cite{sarri13,sarri15,sarri03}, but the equilibrium deformations are very close 
to each other changing at most by a few percent.

After the HF+BCS calculation is performed, we introduce separable spin-isospin 
residual interactions and solve the QRPA equations in the deformed 
ground-states to get GT strength distributions and $2\nu\beta\beta$-decay NMEs. 
The residual force has both particle-hole ($ph$) and particle-particle ($pp$) 
components. 
The repulsive $ph$ force determines to a large extent the structure of the GT
resonance and its location. Its coupling constant $\chi_{ph}^{GT}$ is usually taken 
to reproduce them \cite{sarriguren,sarri_pp,moller,hir,homma}.  We use 
$\chi_{ph}^{GT}=3.0/A^{0.7}{\rm MeV}$. The attractive $pp$ part is basically a 
proton-neutron pairing interaction. We also use a separable form \cite{sarri_pp,hir}
with a coupling constant $\kappa_{pp}^{GT}$ usually fitted to reproduce the experimental
half-lives \cite{homma}. We use in most of this work a fixed value
$\kappa_{pp}^{GT}=0.05$ MeV, although we will explore the dependence of the  
$2\nu\beta\beta$ NMEs on $\kappa_{pp}^{GT}$ in the next section. Earlier studies 
on $^{150}$Nd and $^{150}$Sm carried out in Refs. \cite{yousef09,fang10} using a 
deformed QRPA formalism showed that the results obtained from realistic 
nucleon-nucleon residual interactions based on the Brueckner $G$ matrix for the 
CD-Bonn force produce results in agreement with those obtained from schematic 
separable forces similar to those used here.

The QRPA equations are solved following the lines described in
Refs.~\cite{hir,sarriguren,sarri_pp}. The method we use is as follows.
We first introduce the proton-neutron QRPA phonon operator

\begin{equation}
\Gamma _{\omega _{K}}^{+}=\sum_{\pi\nu}\left[ X_{\pi\nu}^{\omega _{K}}
\alpha _{\nu}^{+}\alpha _{\bar{\pi}}^{+}+Y_{\pi\nu}^{\omega _{K}}
\alpha _{\bar{\nu}} \alpha _{\pi}\right]\, ,  \label{phon}
\end{equation}
where $\alpha ^{+}$ and $\alpha $ are quasiparticle creation and annihilation 
operators, respectively. $\omega _{K}$ labels the RPA excited state and its 
corresponding excitation energy, and $X_{\pi\nu}^{\omega _{K}},Y_{\pi\nu}^{\omega _{K}}$ 
are the forward and backward phonon amplitudes, respectively. 
The solution of the QRPA equations are obtained by solving first a  
dispersion relation \cite{hir,sarri_pp}, which is of fourth order in the excitation  
energies $\omega_K$. The GT transition amplitudes connecting 
the QRPA ground state $\left| 0\right\rangle$ 
($\Gamma _{\omega _{K}} \left| 0 \right\rangle =0$)
to one phonon states $\left| \omega _K \right\rangle$ 
($ \Gamma ^+ _{\omega _{K}} \left| 0 \right\rangle = \left| \omega _K \right\rangle $)
are given in the intrinsic frame by

\begin{equation}
\left\langle \omega _K | \sigma _K t^{\pm} | 0 \right\rangle = 
\mp M^{\omega _K}_\pm \, ,
\end{equation}
where

\begin{eqnarray}
M_{-}^{\omega _{K}}&=&\sum_{\pi\nu}\left( 
v_{\nu}u_{\pi} X_{\pi\nu}^{\omega _{K}}+
u_{\nu}v_{\pi} Y_{\pi\nu}^{\omega _{K}}\right) 
\left\langle \nu\left| \sigma _{K}\right| 
\pi\right\rangle , \\ 
M_{+}^{\omega _{K}}&=&\sum_{\pi\nu}\left( 
u_{\nu}v_{\pi} X_{\pi\nu}^{\omega _{K}}+ 
v_{\nu}u_{\pi} Y_{\pi\nu}^{\omega _{K}}\right) 
\left\langle \nu\left| \sigma _{K}\right| 
\pi\right\rangle \, .
\end{eqnarray}
$v_{\nu,\pi}$ ($u^2_{\nu,\pi}=1-v^2_{\nu,\pi}$) are the BCS occupation amplitudes for 
neutrons and protons. Once the intrinsic amplitudes are calculated, the GT strength 
$B$(GT) in the laboratory frame for a transition 
$I_i K_i (0^+0)\rightarrow I_fK_f(1^+K)$ can be obtained as

\begin{eqnarray}
B_{\omega}({\rm GT}^\pm ) &=& \sum_{\omega_{K}} \left[ \left\langle \omega_{K=0} 
\left| \sigma_0t^\pm \right| 0 \right\rangle ^2 \delta (\omega_{K=0}-
\omega ) \right. \nonumber \\
&& + 2\left. \left\langle \omega_{K=1} \left| \sigma_1t^\pm \right| 
0 \right\rangle ^2 \delta (\omega_{K=1}-\omega ) \right] \, .
\label{bgt}
\end{eqnarray}
To obtain this expression we have used the Bohr and Mottelson factorization \cite{bm}
to express the initial and final nuclear states in the laboratory system in terms 
of the intrinsic states. A quenching factor, $q=g_{A}/ g_{A,{\rm bare}} = 0.79$,
is applied to the weak axial-vector coupling constant and included 
in the calculations. The physical reasons for this quenching have been studied 
elsewhere \cite{caurier05,osterfeld92,bertsch82} and are related to the role of 
non-nucleonic degrees of freedom, absent in the usual theoretical models, and to 
the limitations of model space, many-nucleon configurations, and deep correlations 
missing in these calculations. The implications of this quenching on the description 
of single-$\beta$ and double-$\beta$-decay observables have been considered in
several works \cite{suhonen14,suhonen13,pirinen,fogli08,barea13,caurier12}, where
both the effective value of $g_A$ and the coupling strength of the 
residual interaction in the $pp$ channel are considered free parameters 
of the calculation. It is found that very strong quenching values are needed 
to reproduce simultaneously the observations corresponding to the $2\nu\beta\beta$ 
half-lives and to the single-$\beta$ decay branches. One should note however, that 
the QRPA calculations that require a strong quenching to fit the $2\nu\beta\beta$ 
NMEs were performed within a spherical formalism neglecting possible effects from 
deformation degrees of freedom. Because the main effect of deformation is a reduction 
of the NMEs, deformed QRPA calculations shall demand less quenching to fit the 
experiment.

\begin{table*}[ht]
\label{tablenme}
\begin{center}
\caption{Experimental $2\nu\beta\beta$-decay half-lives $T_{1/2}^{2\nu\beta\beta}$  
from Ref.~\cite{barabash15}, phase-space factors $ G^{2\nu\beta\beta}$ from 
Ref.~\cite{kotila12}, and NMEs extracted from Eq.~(\ref{half-life}) taking bare 
$g_{A,{\rm bare}}=1.273$ and quenched $g_A=1$ factors.}
\vskip 0.5cm
\begin{tabular}{llccccc}
\hline \hline \\
&& $^{76}$Ge & $^{116}$Cd & $^{128}$Te & $^{130}$Te & $^{150}$Nd \\ 
\hline \\
$ T_{1/2}^{2\nu\beta\beta}$ ($10^{21}$ yr) && $ \quad 1.65 \pm 0.14 \quad $ &
$ \quad  0.0287 \pm 0.0013  \quad $ & $ \quad 2000 \pm 300 \quad $ &
$ \quad 0.69 \pm 0.13 \quad $ & $ \quad 0.0082 \pm 0.0009 \quad $ \\
$ G^{2\nu\beta\beta}$ ($10^{-21}$ yr$^{-1}$) && 48.17 & 2764 & 
0.2688 & 1529 & 36430 \\
\\
 & \vline \quad $g_A=1.273$ & 0.136 & 0.136 & 0.052 & 0.037 & 0.070 \\
$M_{GT}^{2\nu\beta\beta}$ (MeV$^{-1}$) &\vline &&&&& \\
& \vline \quad $g_A=1$ & 0.220 & 0.220 & 0.084 & 0.060 & 0.113 \\ \\
\hline\hline
\end{tabular}
\end{center}
\end{table*}

Concerning  the $2\nu\beta\beta$-decay NMEs, the basic expressions for this process, 
within the deformed QRPA formalism used in this work, can be found in Refs. 
\cite{alvarez04,simkovic04,moreno09}. Deformation effects on the $2\nu\beta\beta$ 
NMEs have also been studied within the Projected Hartree-Fock-Bogoliubov model 
\cite{singh07}. Attempts to describe deformation effects on the $0\nu\beta\beta$ 
decay within QRPA models can also be found in Refs. \cite{fang10r,mustonen13}.

The half-life of the $2\nu\beta\beta$ decay can be written as 

\begin{equation}\label{half-life}
\left[ T_{1/2}^{2\nu\beta\beta}\left( 0^+_{\rm gs} \to 0^+_{\rm gs}  
\right) \right] ^{-1}= (g_A)^4\ G^{2\nu\beta\beta}\ \left| (m_ec^2)
M_{GT}^{2\nu\beta\beta}\right| ^2 \, ,
\end{equation}
where $G^{2\nu\beta\beta}$ are the phase-space integrals \cite{kotila12,ahep2016} and 
$M_{GT}^{2\nu\beta\beta}$ the nuclear matrix elements containing the nuclear 
structure part involved in the $2\nu\beta\beta$ process,

\begin{eqnarray}\label{mgt}
&& M_{GT}^{2\nu\beta\beta}=\sum_{K=0,\pm 1}\sum_{m_i,m_f} (-1)^K 
\frac{\langle \omega_{K,m_f}  | 
\omega_{K,m_i}  \rangle } {D} \nonumber \\
&& \times \langle 0_f| \sigma_{-K}t^-| \omega_{K,m_f}  \rangle \: 
\langle \omega_{K,m_i}  | \sigma_Kt^- | 0_i \rangle \, .
\end{eqnarray}
In this equation $|\omega_{K,m_i}\rangle (|\omega_{K,m_f}\rangle)$ are the QRPA 
intermediate $1^+$ states reached from the initial (final) nucleus. $m_i$, $m_f$ 
are labels that classify the intermediate $1^+$ states that are reached from
different initial $| 0_i \rangle$ and final $ | 0_f \rangle$ ground states.
The overlaps $\langle \omega_{K,m_f}  | \omega_{K,m_i}  \rangle $ 
take into account the non-orthogonality of the intermediate states. Their 
expressions can be found in Ref.~\cite{simkovic04}. The energy 
denominator $D$ involves the energy of the emitted leptons, which is given on 
average by $\frac{1}{2} Q_{\beta \beta}+m_e$, as well as the excitation energies of 
the intermediate nucleus. In terms of the QRPA excitation energies the denominator
can be written as 
\begin{equation}\label{den1}
D_1= \frac{1}{2} (\omega_K^{m_f}+ \omega_K^{m_i}),
\end{equation}
where  $\omega_K^{m_i} (\omega_K^{m_f})$ is the QRPA excitation energy relative to the 
initial (final) nucleus. It turns out that the NMEs are quite sensitive to the 
values of the denominator, especially for low-lying states, where the denominator
takes smaller values. Thus, it is a common practice to use some experimental 
normalization of this denominator to improve the accuracy of the NMEs.
In this work we also consider the denominator $D_2$, which is corrected with the 
experimental energy $\bar{\omega}_{1_1^+}$ of the first $1^+$ state in the intermediate 
nucleus relative to the mean ground-state energy of the initial and final nuclei, 
in such a way that the experimental energy of the first $1^+$ state is reproduced
by the calculations,

\begin{equation}\label{den2}
D_2= \frac{1}{2} \left[ \omega_K^{m_f}+ \omega_K^{m_i}-\left( \omega_K^{1_f}+ 
\omega_K^{1_i}\right) \right]  
+\bar{\omega}_{1_1^+} \, .
\end{equation}
Running $2\nu\beta\beta$ sums will be shown later for the two 
choices of the denominator $D_1$ and $D_2$. When the ground state in the 
intermediate nucleus of the double-$\beta$-decay partners is a $1^+$ state, the
energy $\bar{\omega}_{1_1^+}$ is given by 

\begin{equation}\label{wexp}
\bar{\omega}_{1_1^+}=\frac{1}{2}(Q_{EC}+Q_{\beta^-})_{\rm exp} \, ,
\end{equation}
where $Q_{EC}$ and $Q_{\beta^-}$ are the experimental energies of the decays of the 
intermediate nucleus into the parent and daughter partners, respectively.
This is the case of $^{116}$In and $^{128}$I, which are both  $1^+$ ground states.
In the other cases, although the ground state in the intermediate nuclei are not
$1^+$ states, the first $1^+$ excited states appear at a very low excitation energy,
E=0.086 MeV in $^{76}$As \cite{thies12}, E=0.043 MeV in $^{130}$I \cite{puppe12}, and
E=0.11 MeV in $^{150}$Pm \cite{guess11}. Therefore, to a good approximation
we also determine $\bar{\omega}_{1_1^+}$  using Eq. (\ref{wexp}).

The existing measurements for the $2\nu\beta\beta$-decay half-lives 
($T_{1/2}^{2\nu\beta\beta}$) have been recently analyzed in Ref.~\cite{barabash15}. 
Adopted values for such half-lives can be seen in Table I.
Using the phase-space factors from the evaluation \cite{kotila12} that involves 
exact Dirac wave functions including  electron screening and finite nuclear size 
effects, we obtain the experimental NMEs shown in Table I, for
bare $g_{A,{\rm bare}}=1.273$ and quenched $g_A=1$ factors.
It should be clear that the theoretical NMEs defined in Eq. (\ref{mgt})
do not depend on the $g_A$ factors. Hence, the value obtained for the experimental
NMEs extracted from the experimental half-lives through Eq. (\ref{half-life}) depend 
on the $g_A$ value used in this equation.

\section{Results}
\label{sec:results}

\subsection{Gamow-Teller strength distributions}

\begin{figure}[htb]
\begin{center}
\includegraphics[width=8.5cm]{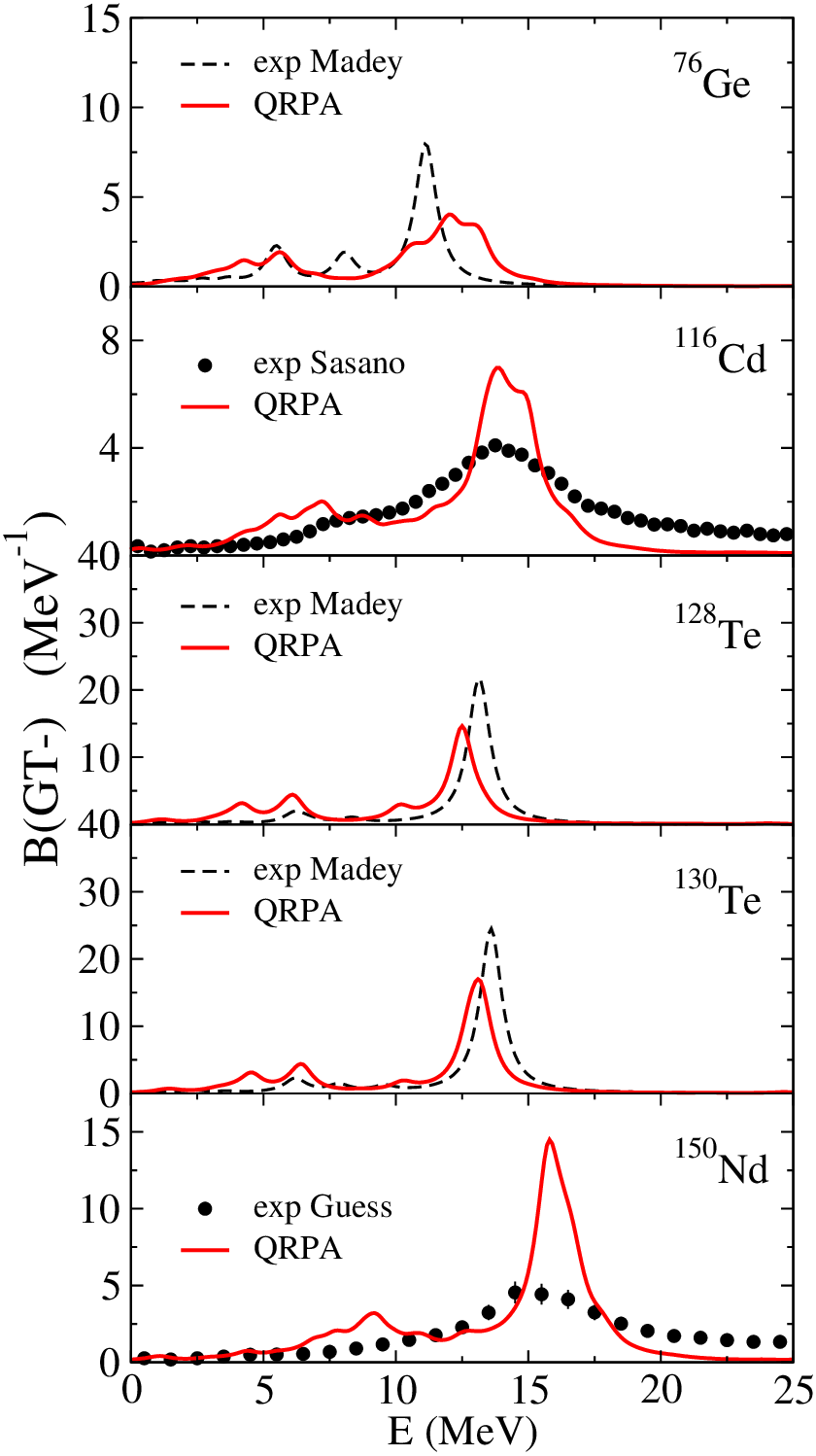}
\caption{Experimental $B({\rm GT}^-)$ from CERs \cite{madey89,sasano12,guess11}
in $^{76}$Ge, $^{116}$Cd, $^{128}$Te, $^{130}$Te, and $^{150}$Nd plotted versus the 
excitation energy of the daughter nuclei are compared with folded SLy4-QRPA 
calculations (see text).}
\label{fig1}
\end{center}
\end{figure}

\begin{figure}[htb]
\begin{center}
\includegraphics[width=8.5cm]{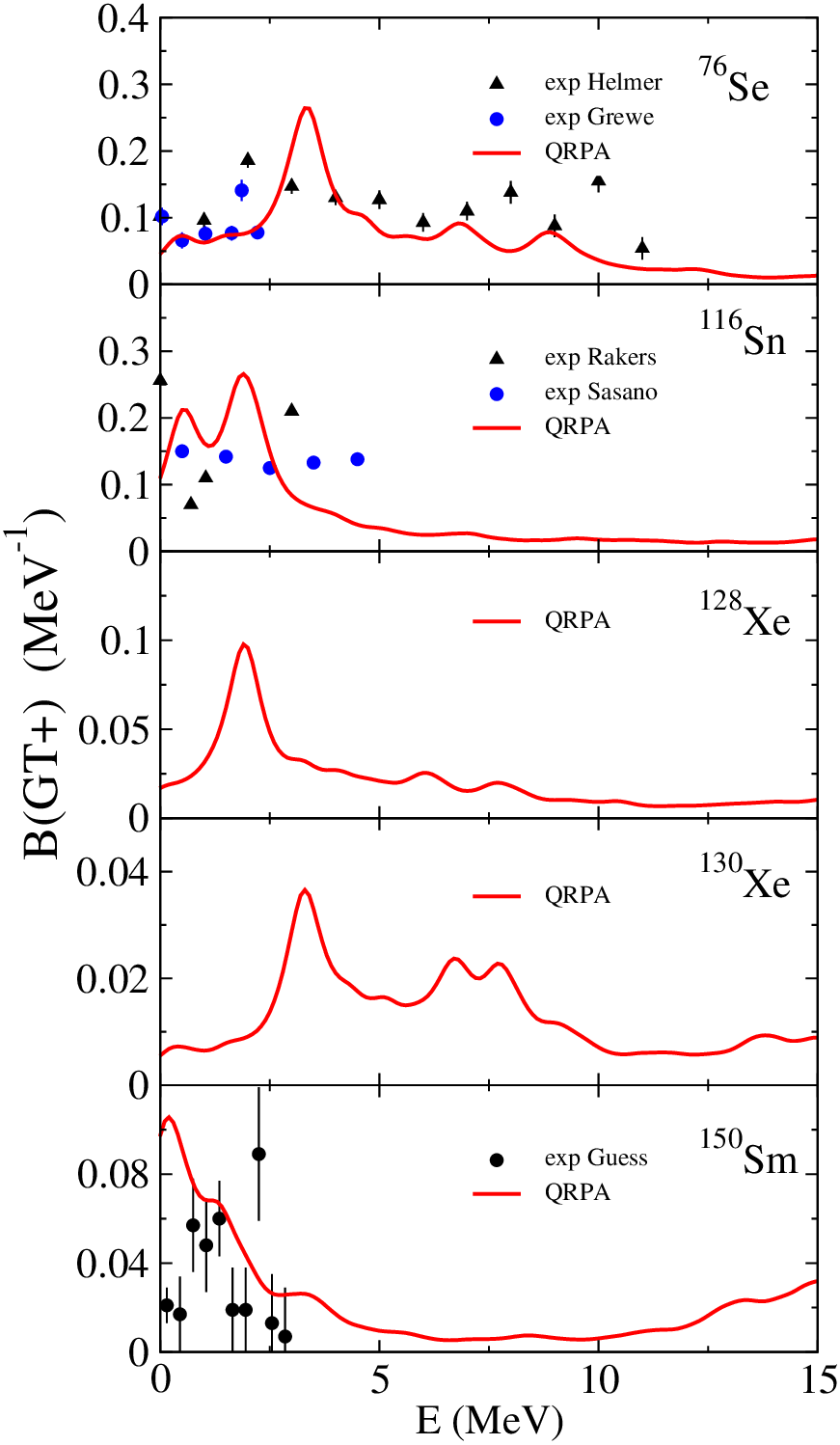}
\caption{Same as in Fig. \ref{fig1}, but for the $B({\rm GT}^+)$ in $^{76}$Se, 
$^{116}$Sn, $^{128}$Xe, $^{130}$Xe, and $^{150}$Sm. Experimental data are from 
CERs \cite{helmer97,grewe08,rakers05,sasano12,guess11}.}
\label{fig2}
\end{center}
\end{figure}

The energy distributions of the GT strength obtained from our formalism are 
displayed in Figs. \ref{fig1}-\ref{fig2}. Figure \ref{fig1} contains the 
$B({\rm GT}^-)$ strength distributions for $^{76}$Ge, $^{116}$Cd, $^{128}$Te, 
$^{130}$Te, and $^{150}$Nd. The theoretical curves correspond to the calculated 
distributions folded with 1 MeV width Breit-Wigner functions, in such a way that 
the discrete spectra obtained in the calculations appear now as continuous 
curves. They give the GT strength per MeV and the area below the curves in a 
given energy interval gives us directly the GT strength contained in that energy
interval. We compare our QRPA results from 
SLy4 obtained with the selfconsistent deformations with the experimental 
strengths extracted from CERs \cite{madey89,sasano12,guess11}.
In the cases of $^{76}$Ge, $^{128}$Te, and $^{130}$Te, the data from \cite{madey89} 
includes the total GT measured strength of the resonances and their energy 
location. Namely, $B$(GT)=12.43 at E=11.13 MeV in  $^{76}$Ge,
$B$(GT)=34.24 at E=13.14 MeV in  $^{128}$Te,
$B$(GT)=38.46 at E=13.59 MeV in  $^{130}$Te. We have folded these strengths with 
the same functions used for the calculations to facilitate the comparison. They
can be seen with dashed lines in Fig.\ref{fig1}.

\begin{figure}[htb]
\begin{center}
\includegraphics[width=8.5cm]{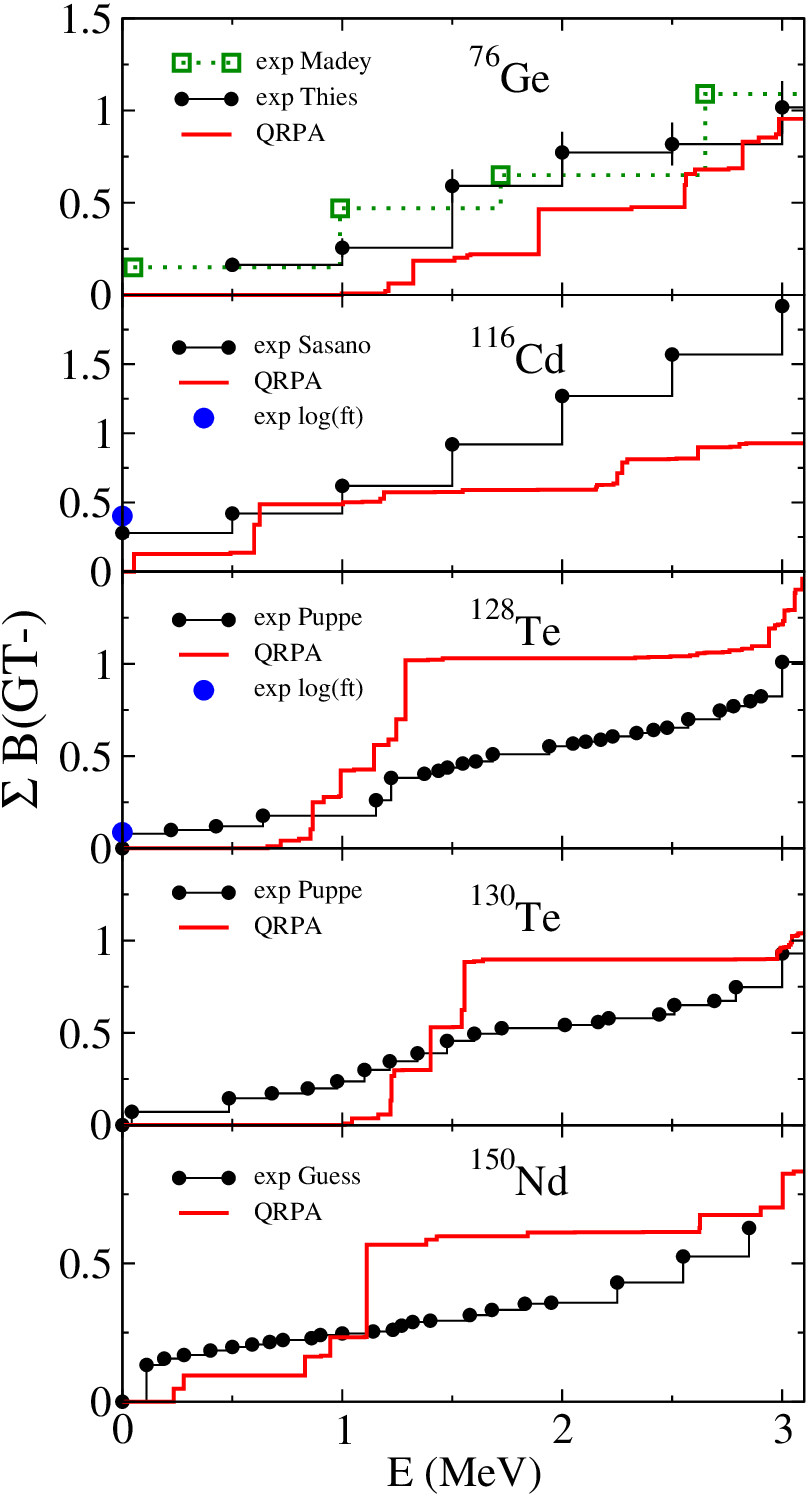}
\caption{Accumulated GT strength $B({\rm GT}^-)$ in the low-energy range.
SLy4-QRPA calculations are compared with data from Refs.
\cite{madey89,guess11,thies12,sasano12,puppe12}. Also shown in $^{116}$Cd
and $^{128}$Te are the $B({\rm GT}^-)$ values extracted from the experimental
electron captures on the intermediate nuclei $^{116}$I and $^{128}$In.}
\label{fig3}
\end{center}
\end{figure}

\begin{figure}[htb]
\begin{center}
\includegraphics[width=8.5cm]{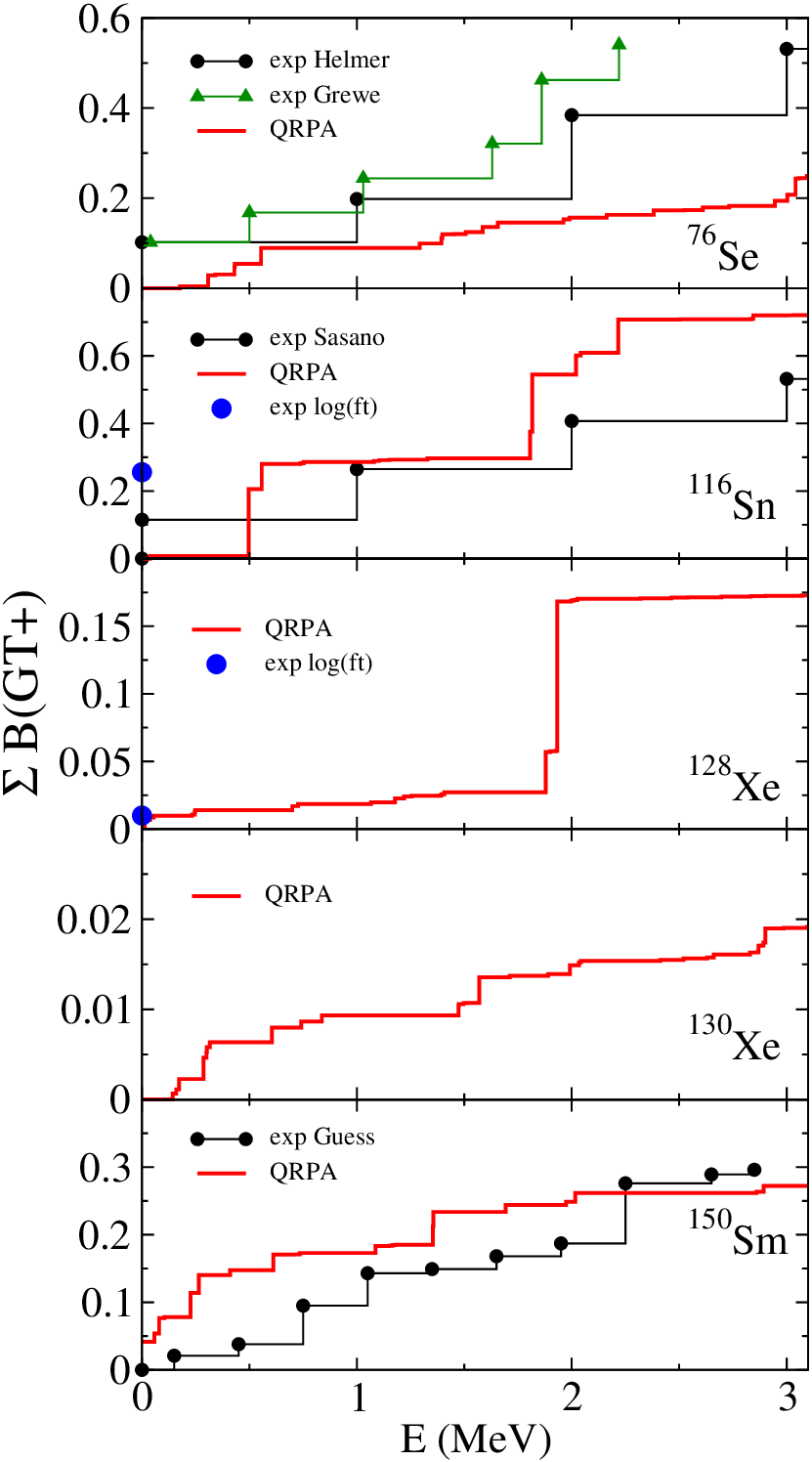}
\caption{Same as in Fig. \ref{fig2}, but plotted as accumulated strength in the 
low-energy range. Also shown in $^{116}$Sn and $^{128}$Xe are the $B({\rm GT}^+)$ 
values extracted from the experimental $\beta^-$ decay of the intermediate nuclei 
$^{116}$I and $^{128}$In.}
\label{fig4}
\end{center}
\end{figure}

Figure \ref{fig2} 
contains the $B({\rm GT}^+)$ strength distributions corresponding to $^{76}$Se, 
$^{116}$Sn, $^{128}$Xe, $^{130}$Xe, and $^{150}$Sm. The QRPA results 
folded with the same 1 MeV width Breit-Wigner functions
are compared with the experimental strengths extracted from CERs 
\cite{helmer97,grewe08,rakers05,sasano12,guess11}.
On the other hand, Figs. \ref{fig3}-\ref{fig4} contain the accumulated GT strength 
in the low excitation energy. Figure \ref{fig3} contains the same cases as in Fig. 
\ref{fig1} with additional high-resolution data from Ref. \cite{thies12} for 
$^{76}$Ge and from Ref. 
\cite{puppe12} for $^{128,130}$Te. Figure \ref{fig4} contains the same cases as in 
Fig. \ref{fig2}, but as accumulated strengths in the low-energy range.

One should notice that the measured strength extracted from the cross sections 
contains two types of contributions that cannot be disentangled, namely GT 
($\sigma t^\pm$ operator) and isovector spin monopole (IVSM) ($r^2 \sigma t^\pm$ 
operator). Thus, the measured strength corresponds actually to $B$(GT+IVSM).
Different theoretical calculations evaluating the contributions from both GT and 
IVSM modes are available in the literature
\cite{guess11,sasano12,hamamoto00,bes12,civitarese14}. The general conclusion 
tells us that in the (p,n) direction the strength distribution below 20 MeV
is mostly caused by the GT component, although non-negligible contributions from 
IVSM components are found between 10 and 20 MeV. Above 20 MeV, there is no 
significant GT strength in the calculations. In the (n,p) direction 
the GT strength is expected to be strongly Pauli blocked in nuclei with more 
neutrons than protons and therefore, the measured strength is mostly due to the 
IVSM resonance. 
Nevertheless, the strength found in low-lying isolated peaks is associated with 
GT transitions because the continuous tail of the IVSM resonance is very small 
at these energies and is not expected to exhibit any peak. In summary, the 
measured strength in the ($p,n$) direction can be safely assigned to be GT in 
the low energy range below 10 MeV and with some reservations between 10 and 20 MeV. 
Beyond 20 MeV the strength would be practically due to IVSM. On the other hand, 
the measured strength in the ($n,p$) direction would be due to IVSM transitions, 
except in the low-lying excitation energy below 2-3 MeV, where the isolated 
peaks observed can be attributed to GT strength. This is the reason why we
plot experimental data in Fig. \ref{fig4} only up to 3 MeV.

In general terms, we reproduce fairly well the global properties of the GT 
strength distributions, including the location of the GT$^-$ resonance and the 
total strength measured (see Fig. \ref{fig1}). In the (n,p) 
direction, the GT$^+$ strength is strongly suppressed (compare the vertical scales
in Figs. \ref{fig1} and \ref{fig2}). As expected, a strong suppression of
GT$^+$ takes place in nuclei with a large neutron excess. The experimental 
information on  GT$^+$ strengths is mainly limited to the low-energy region 
and it is fairly well reproduced by the 
calculations. The accumulated strengths in the low-energy range shown in Figs. 
\ref{fig3}-\ref{fig4} show more clearly the degree of accuracy achieved by the 
calculations. Although a detailed spectroscopy is beyond the capabilities of our 
model and the isolated transitions are not well reproduced by our calculations,
the overall agreement with the total strength contained in this reduced
energy interval, as well as with the profiles of the accumulated strength
distributions, is satisfactory. In general, the experimental B(GT$^-$) shows 
spectra more fragmented than the calculated ones, but the total strength up to
3 MeV is well reproduced with the only exception of $^{116}$Cd, where we obtain
less strength than observed. The total measured B(GT$^+$) strength up to 3 MeV
is especially well reproduced in the case of $^{150}$Sm, whereas it is somewhat
underestimated in $^{76}$Se  and overestimated in $^{116}$Sn.

We can see in Figs. \ref{fig3} and \ref{fig4} with blue dots the $B$(GT) values
extracted from the decays of the intermediate $1^+$ nuclei $^{116}$In and $^{128}$I.
They can be compared with experimental results extracted from CERs, as well as 
with the theoretical calculations. The electron capture experiment on $^{116}$In 
\cite{wrede13} gives $ft=2.84\times 10^4$ s with a corresponding strength 
B(GT$^-$)=0.402. The $\beta^-$ decay yields B(GT$^-$)=0.256 \cite{rakers05}. The 
decay of $^{128}$I yields  B(GT$^-$)=0.087 and B(GT$^+$)=0.079 \cite{puppe12}.
The sensitivity of these distributions to the effective interactions and to
nuclear deformation was discussed in previous works 
\cite{alvarez04,sarri03,sarri12,sarri15,moreno09}.
Different calculations \cite{simkovic04,yousef09,civitarese14,suhonen14,delion15}
based also on QRPA formalisms with different degrees of sophistication agree 
qualitatively in the description of the single $\beta$ branches of 
double-$\beta$-decay partners.

\begin{figure}[htb]
\begin{center}
\includegraphics[width=8.5cm]{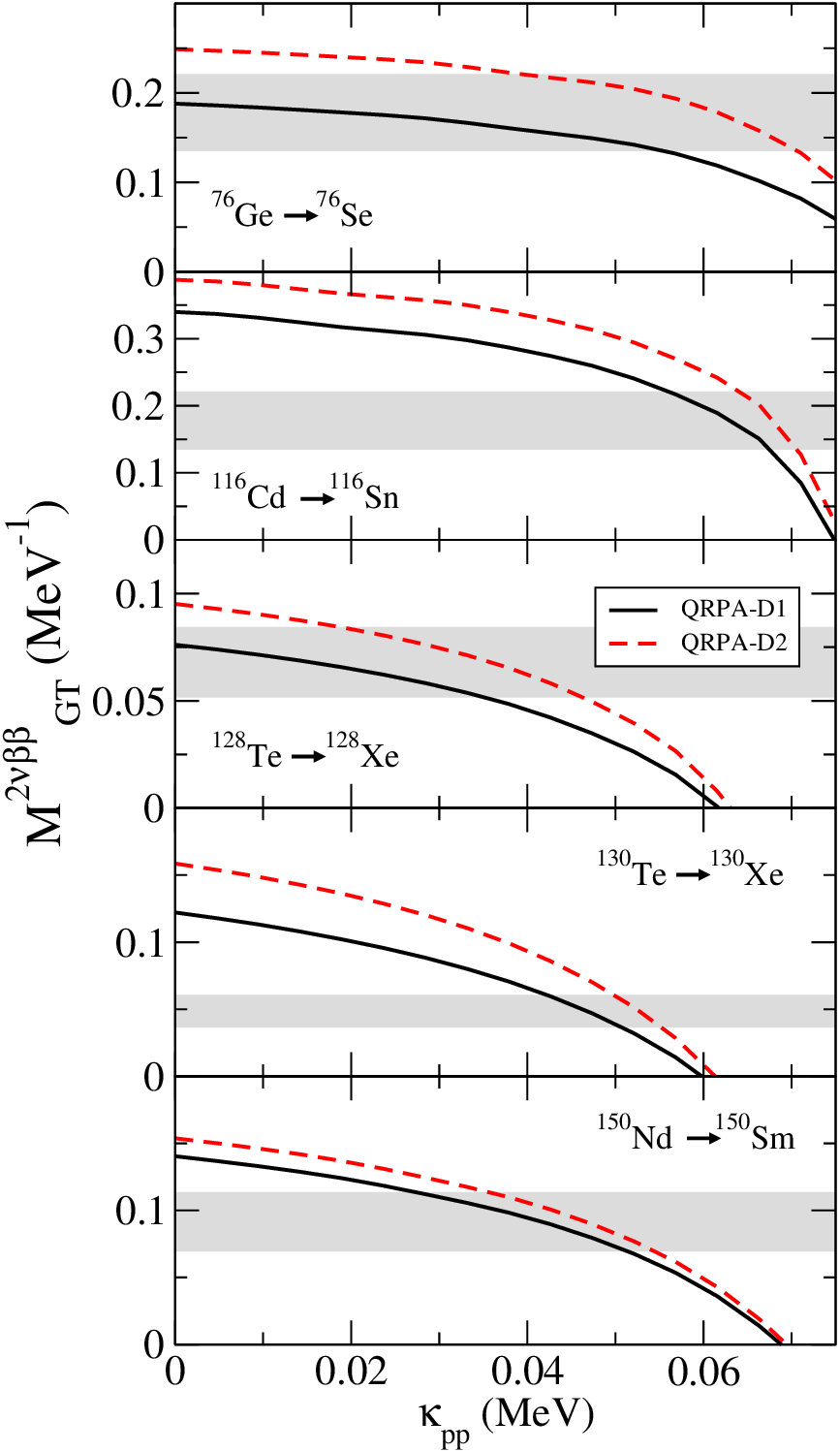}
\caption{
Nuclear matrix element for the $2\nu\beta\beta$ decay 
of $^{76}$Ge, $^{116}$Cd, $^{128}$Te, $^{130}$Te, and $^{150}$Nd 
as a function of the coupling strength $\kappa_{pp}^{GT}$.
Solid lines correspond to calculations with the energy denominator
$D_1$, while dashed lines correspond to $D_2$.
The gray area corresponds to the NME experimental range obtained from the 
measured half-lives using bare $g_A=1.273$ and quenched $g_A=1$ factors.}
\label{fig5}
\end{center}
\end{figure}

\begin{figure}[htb]
\begin{center}
\includegraphics[width=8.5cm]{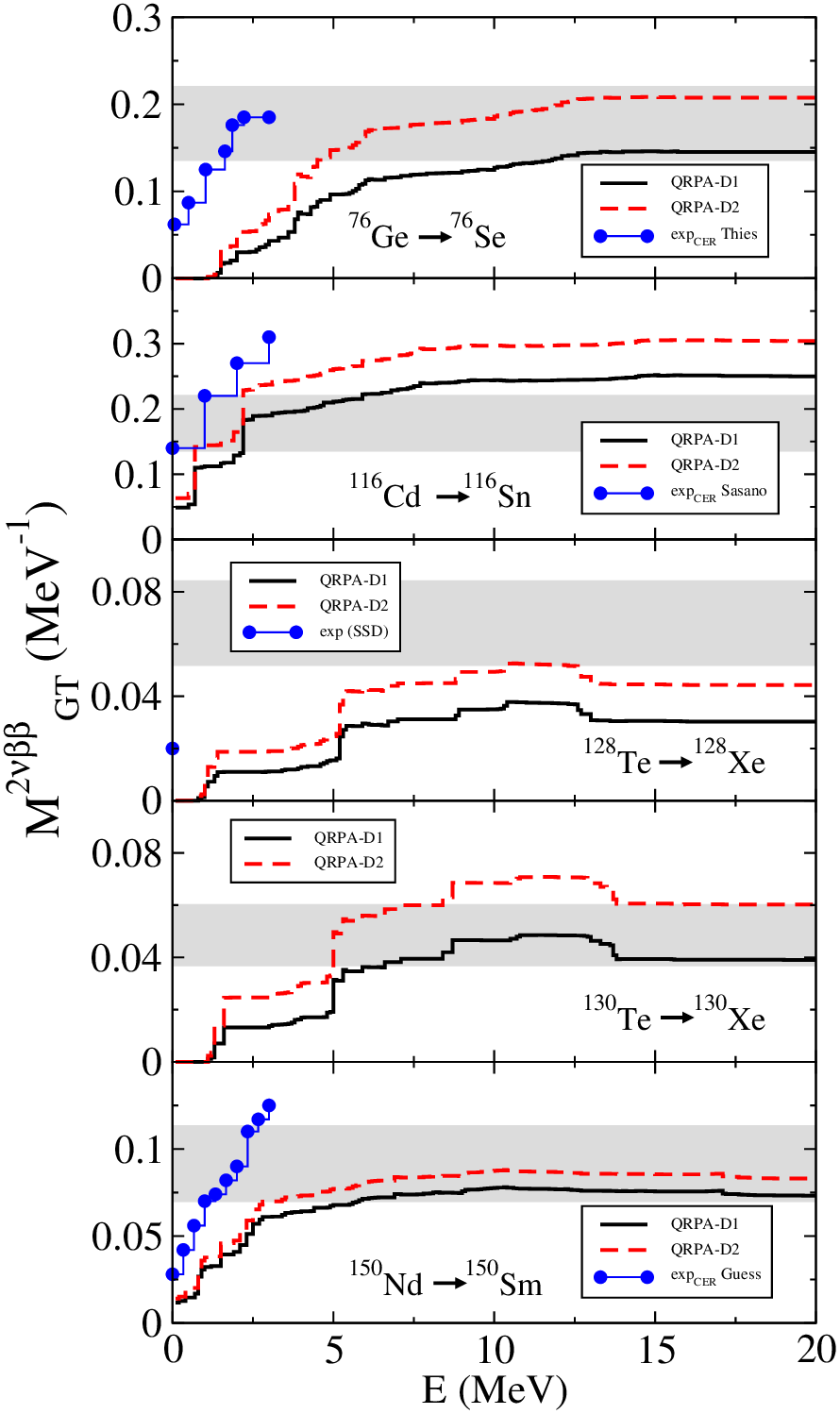}
\caption{Running sums of the $2\nu\beta\beta$ NME in $^{76}$Ge, $^{116}$Cd, 
$^{128}$Te, $^{130}$Te, and $^{150}$Nd as a function of the excitation energy in 
the intermediate nucleus. Solid and dashed lines and shaded areas are as in 
Fig. \ref{fig5}. See text.}
\label{fig6}
\end{center}
\end{figure}

\subsection{Double-$\beta$ decay}
\label{dbd}

It is well known that the $2\nu\beta\beta$ NMEs are very sensitive to the residual 
interactions, as well as to differences in deformation between initial and final 
nuclei \cite{alvarez04,simkovic04}. We show in Fig. \ref{fig5} the NMEs calculated 
with the selfconsistent deformations as a function of the $pp$ coupling constant
of the residual force for the decays of $^{76}$Ge, $^{116}$Cd, $^{128}$Te, $^{130}$Te, 
and $^{150}$Nd. The shaded bands correspond to the experimental NMEs extracted 
from the measured $2\nu\beta\beta$ half-lives. For each nucleus the band is delimited
by the lower and upper values obtained using 
bare ($g_A=1.273$) and quenched values, respectively, (see Table I). 
Results obtained with the energy denominator $D_1$ are displayed with solid 
lines, whereas results obtained with $D_2$ are shown with dashed lines. $D_2$ 
denominators produce in all cases larger NMEs than  $D_1$. We can see that the 
experimental NMEs contained in the shaded region are reproduced within some windows 
of the parameter $\kappa_{pp}^{GT}$. It is not our purpose here to get the best fit 
or the optimum value of $\kappa_{pp}^{GT}$ that reproduces the experimental NMEs 
because this value will change by changing $\chi_{ph}^{GT}$ or the underlying mean 
field structure. In this work we take $\kappa_{pp}^{GT}=0.05$ MeV as an approximate 
value that reproduces reasonably well the experimental information on both single 
$\beta$ branches and $2\nu\beta\beta$ NMEs. 

Figure \ref{fig6} shows the running sums for the $2\nu\beta\beta$ NMEs calculated
with $\kappa_{pp}^{GT}=0.05$ MeV. These are the partial contributions to the NMEs 
of all the $1^+$ states in the intermediate nucleus up to a given energy. 
Obviously, the final values reached by the calculations at 20 MeV in Fig. \ref{fig6} 
correspond to the values in Fig. \ref{fig5} at $\kappa_{pp}^{GT}=0.05$ MeV.
The final values of the running sums for other  $\kappa_{pp}^{GT}$ can be estimated 
by looking at the corresponding  $\kappa_{pp}^{GT}$ values in Fig. \ref{fig5}.
As in the previous figure, we also show the results 
obtained with denominators $D_1$ (solid) and $D_2$ (dashed). The main difference 
between them is originated at low excitation energies, where the relative effect 
of using shifted energies is enhanced. The effect at larger energies is negligible 
and we get a constant difference between $D_1$ and $D_2$, which is the difference 
accumulated in the first few MeV. The contribution to the $2\nu\beta\beta$ 
NMEs in the region between 10-15 MeV that can be seen in most cases, is due to
the GT resonances observed in Fig. \ref{fig1}. This contribution is small because
the joint effects of large 
energy denominators in Eq. (\ref{mgt}) and the mismatch between the excitation
energies of the GT$^-$ and GT$^+$ resonances.

The running sums are very useful to discuss the extent to which the 
single-state-dominance hypothesis applies. This hypothesis tells us that, to a 
large extent, the $2\nu\beta\beta$ NMEs will be given by the transition through 
the ground state of the intermediate odd-odd nucleus in those cases where this 
ground state is a $1^+$ state reachable by allowed GT transitions.
One important consequence of the SSD hypothesis would be that the  
half-lives for $2\nu\beta\beta$ decay could be extracted accurately from simple
experiments, such as single $\beta^-$ and electron capture measurements of the 
intermediate nuclei to the $0^+$ ground states of the neighbor even-even nuclei.
Theoretically, the SSD hypothesis would also imply an important simplification
of the calculations because to describe the  $2\nu\beta\beta$ decay from ground
state to ground state, only the wave function of the $1^+$ ground state of the
intermediate nucleus would be needed. 
Because not all of the 
double-$\beta$ decaying nuclei have $1^+$ ground states in the intermediate nuclei 
(only $^{116}$In and $^{128}$I in the nuclei considered here), the SSD condition is
extended by considering the relative contributions of the low-lying excited states 
in the intermediate nuclei to the total $2\nu\beta\beta$ NMEs. This is called 
low-lying-single-state dominance \cite{moreno09} and can be studied in all 
$2\nu\beta\beta$ nuclei. 
From the results displayed in Fig. \ref{fig6} we cannot establish clear evidences 
for SSD hypothesis from our calculations. Nevertheless, it is also worth mentioning
that our NMEs calculated up to 5 MeV, already account for most of the total NME
calculated up to 20 MeV. This results agrees qualitatively with other results 
obtained in different QRPA calculations \cite{civi_suho_98,simk_ssd}. 

The SSD hypothesis can be tested experimentally in the decays of $^{116}$Cd 
and $^{128}$Te where the intermediate nuclei have $1^+$ ground states. By 
measuring the two decay branches of $^{116}$In and $^{128}$I, the $log(ft)$ values
of the ground state to ground state ($1^+ \rightarrow 0^+$) can be extracted.
From these values one can obtain the GT strength, 
 
\begin{equation}
B({\rm GT})=\frac{3A}{g_A^2 ft}\ ,
\end{equation}
with  A=6289 s \cite{hardy15}.
Finally the $2\nu\beta\beta$ NME within SSD is evaluated as
\begin{eqnarray}
&&M_{GT}^{2\nu\beta\beta}(SSD)=\frac{\left[ B({\rm GT}^-)B({\rm GT}^+) \right] ^{1/2}}
{(Q_{\beta^-} + Q_{\rm EC})/2} \nonumber \\
&&= \frac{6A}{\left[ ft_{\rm EC}\right] ^{1/2} \left[ ft_{\beta^-}\right] ^{1/2} 
g_A^2(Q_{\beta^-} + Q_{\rm EC})}\, .
\label{dgt}
\end{eqnarray}

One can also determine the $2\nu\beta\beta$ NME running sums using 
the experimental $B$(GT) extracted from CERs and using the same phases for the 
matrix elements if one can establish a one-to-one correspondence between the 
intermediate states reached from parent and daughter. Then, one can construct 
the $2\nu\beta\beta$ NMEs from the measured GT strengths and energies in the 
CERs in the parent and daughter partners,

\begin{equation}
M_{GT}^{2\nu\beta\beta}(LLSD)=\sum_m \frac{\left[ B_m({\rm GT}^+)B_m({\rm GT}^-) 
\right] ^{1/2}} {E_m+(Q_{\beta^-} + Q_{\rm EC})/2}\, , 
\label{dgtlls}
\end{equation}
where $E_m$ is the excitation energy of the $m$th $1^+$ state relative to
the ground state of the intermediate nucleus. Experimental $2\nu\beta\beta$ 
NMEs running sums have been determined along this line using experimental 
$B({\rm GT})$ from CERs in Ref. \cite{thies12} for $^{76}$Ge, in 
Ref. \cite{sasano12} for $^{116}$Cd, and in Ref. \cite{guess11} for $^{150}$Nd.
In the case of $^{128,130}$Te they have not been determined because of the lack of 
data in the (n,p) direction. They can be seen in Fig. \ref{fig6} under the label 
exp$_{CER}$.

In the case of $^{76}$Ge, the $2\nu\beta\beta$ NMEs are constructed by combining 
the GT$^-$ data from $^{76}$Ge($^3$He,t)$^{76}$As \cite{thies12} with those for 
GT$^+$ transitions from $^{76}$Se(d,$^2$He)$^{76}$As \cite{grewe08}. A large 
fragmentation of the GT$^−$ strength was found in the experiment, not only at
high excitation energies, but also at low excitation energy, which is rather unusual. 
In addition, a lack of correlation between the GT excitation energies from the 
two different branches was 
also observed. Thus, for the evaluation of the $2\nu\beta\beta$ NMEs a one-to-one 
connection between the $B({\rm GT}^-$) and $B({\rm GT}^+$) transitions leading to 
the excited state in the intermediate nucleus needs to be established. In 
particular, since the spectra from the two CER experiments had rather different 
energy resolutions, the strength was accumulated in similar bins to evaluate the 
$2\nu\beta\beta$ NMEs \cite{thies12}. The summed matrix element amounted to 
0.186  MeV$^{-1}$ up to an excitation energy of 2.22 MeV.

In the case of $^{116}$Cd, $^{116}$Cd(p,n)$^{116}$In and $^{116}$Sn(n,p)$^{116}$In 
\cite{sasano12} CERs were used to evaluate the LLSD $2\nu\beta\beta$ NMEs. 
The running sum starts at 0.14 MeV$^{-1}$ at zero excitation energy and reaches 
a value of 0.31 MeV$^{-1}$ at 3 MeV  excitation energy. The value at zero energy 
can be compared with the value obtained by using the $ft$-values of the decay 
in $^{116}$In mentioned above. The value constructed in this way amounts to 
NME(SSD)=0.168 MeV$^{-1}$ \cite{wrede13}.
In the case of $^{128}$Te and $^{130}$Te the lack of experimental information in
the GT$^+$ direction prevents us from evaluating the experimental LLSD estimates.
However, an estimate of $M_{GT}^{2\nu\beta\beta}$(SSD)=0.019 MeV$^{-1}$ in $^{128}$Te 
can be obtained from the $log(ft)$ values of the decay in $^{128}$I. Finally, in 
the case of $^{150}$Nd, although the intermediate nucleus $^{150}$Pm is not a $1^+$ 
state, assuming that the excited $1^+$ state at 0.11 MeV excitation energy observed in 
$^{150}$Nd$(^3$He,$t)^{150}$Pm corresponds to all the GT strength measured between
50 keV and 250 keV in the reaction $^{150}$Sm$(t,^3$He)$^{150}$Pm, one obtains an 
estimate for the SSD $M_{GT}^{2\nu\beta\beta}$(SSD)=0.028 MeV$^{-1}$ \cite{guess11}. 
Extending the running sum by associating the corresponding GT strengths bins
from the reactions in 
both directions and assuming a coherent addition of all the bins,
one gets $M_{GT}^{2\nu\beta\beta}$(SSD)=0.13 MeV$^{-1}$ \cite{guess11} up to an 
excitation energy in the intermediate nucleus of 3 MeV. This experimental running 
sum is included in Fig. \ref{fig6}. In all the cases the experimental running sum
is larger than the calculations and tend to be larger than the experimental values 
extracted from the half-lives. However, one should always keep in mind that the 
present experimental LLSD estimates are indeed upper limits because the 
phases of the NMEs are considered always positive. Although the present calculations
favor coherent phases in the low-energy region, the phases could change depending
on the theoretical model. In particular the sensitivity of these phases to the
$pp$ residual interaction has been studied in Ref. \cite{fang10}.

\section{Summary and Conclusions}
\label{sec:conclusions}

In summary, using a theoretical approach based on a deformed HF+BCS+QRPA
calculation with effective Skyrme interactions, pairing correlations, 
and spin-isospin residual separable forces in the $ph$ and $pp$ channels,
we have studied simultaneously the GT strength distributions of the
double-$\beta$-decay partners  ($^{76}$Ge, $^{76}$Se), ($^{116}$Cd, $^{116}$Sn), 
($^{128}$Te, $^{128}$Xe), ($^{130}$Te, $^{130}$Xe), and ($^{150}$Nd, $^{150}$Sm) 
reaching the intermediate nuclei $^{76}$As, $^{116}$In, $^{128}$I, $^{130}$I, and 
$^{150}$Pm, respectively, as well as their  $2\nu\beta\beta$ NMEs. In this work 
we use reasonable choices for the two-body effective interaction, residual 
interactions, deformations, and quenching factors. The sensitivity of the results 
to the various ingredients in the theoretical model was discussed elsewhere.

Our results for the energy distributions of the GT strength
have been compared with recent data from CERs,
whereas the calculated $2\nu\beta\beta$ NMEs have been compared with 
the experimental values extracted from the measured half-lives for 
these processes, as well as with the running sums extracted from CERs

The theoretical approach used in this work has demonstrated to be well suited to 
account for the rich variety of experimental information available on the nuclear 
GT response. The global properties of the energy distributions of the GT strength
and the $2\nu\beta\beta$ NMEs are well reproduced, with the exception of a 
detailed description of the low-lying GT strength distributions that could
clearly be improved. The  $2\nu\beta\beta$ NMEs extracted from the experimental
half-lives are also reproduced by the calculations with some overestimation
(underestimation) in the case of $^{116}$Cd ($^{128}$Te). 

We have also upgraded the theoretical analysis of SSD and LLSD hypotheses 
and we have compared our calculations with the experimental running sums 
obtained by considering recent measurements from CERs  and decays of the 
intermediate nuclei.

It will be interesting in the future to extend these calculations by including
all the double-$\beta$-decay candidates and to explore systematically the
potential of this method. It will be also interesting to explore the consequences 
of the isospin symmetry restoration, as it was investigated in Ref. \cite{isospin}.
In HF+BCS and QRPA neither the ground states nor the excited states are isospin 
eigenstates, but the expectation values of the $T_z$ operator are conserved. This 
implies that in the $B$(GT$^-$) the transition operator connects states with a given 
expectation value of $T_z=(N-Z)/2$ to states with expectation value of $T_z=(N-Z)/2-1$.

\section*{Acknowledgments} 

This work was supported by MINECO (Spain) under Research Grant
No.~FIS2014-51971-P.
O.M. acknowledges support from a Marie Curie International Outgoing Fellowship within 
the European Union Seventh Framework Programme, under Grant Agreement 
PIOF-GA-2011-298364 (ELECTROWEAK).

\end{document}